\input amstex
\documentstyle{amsppt}
%
\catcode`@=11
\redefine\output@{%
  \def\break{\penalty-\@M}\let\par\endgraf
  \ifodd\pageno\global\hoffset=105pt\else\global\hoffset=8pt\fi  
  \shipout\vbox{%
    \ifplain@
      \let\makeheadline\relax \let\makefootline\relax
    \else
      \iffirstpage@ \global\firstpage@false
        \let\rightheadline\frheadline
        \let\leftheadline\flheadline
      \else
        \ifrunheads@ 
        \else \let\makeheadline\relax
        \fi
      \fi
    \fi
    \makeheadline \pagebody \makefootline}%
  \advancepageno \ifnum\outputpenalty>-\@MM\else\dosupereject\fi
}
\def\Beta{\mathchar"0\hexnumber@\rmfam 42}
\redefine\mm@{2010} 
\catcode`\@=\active
\nopagenumbers
\chardef\textvolna='176

\chardef\bigalpha='013
\def\negskp{\hskip -2pt}

\chardef\degree="5E
\def\compos{\,\raise 1pt\hbox{$\sssize\circ$} \,}

\def\blue#1{#1}

\gdef\darkred#1{#1}
\gdef\darkblue#1{#1}
\gdef\green#1{#1}
\catcode`#=11\def\diez{#}\catcode`#=6
\catcode`&=11\catcode`&=4
\catcode`_=11\def\podcherkivanie{_}\catcode`_=8
\catcode`\^=11\catcode`\^=7
\catcode`~=11\catcode`~=\active
\catcode`\%=11\def\procent{
\def\mycite#1{\cite{\blue{#1}}\immediate\special{ps:
     ShrHPSdict begin /ShrBORDERthickness 0 def}}
\def\myciterange#1#2#3#4{\cite{\blue{#2#3#4}}\immediate\special{ps:
     ShrHPSdict begin /ShrBORDERthickness 0 def}}
\def\mytag#1{%
    \tag#1}
\def\mythetag#1{\thetag{\blue{#1}}\immediate\special{ps:
     ShrHPSdict begin /ShrBORDERthickness 0 def}}
\def\myrefno#1{\no#1}
\def\myhref#1#2{\blue{#2}\immediate\special{ps:
     ShrHPSdict begin /ShrBORDERthickness 0 def}}
\def\myEarXivlink{\myhref{http://arXiv.org}{http:/\negskp/arXiv.org}}
\def\myGeoCities{\myhref{http://www.geocities.com}{GeoCities}}
\def\mytheorem#1{\csname proclaim\endcsname{Theorem #1}}
\def\mytheoremwithtitle#1#2{\csname proclaim\endcsname{Theorem #1#2}}
\def\mythetheorem#1{\blue{#1}\immediate\special{ps:
     ShrHPSdict begin /ShrBORDERthickness 0 def}}
\def\mylemma#1{\csname proclaim\endcsname{Lemma #1}}
\def\mylemmawithtitle#1#2{\csname proclaim\endcsname{Lemma #1#2}}
\def\mythelemma#1{\blue{#1}\immediate\special{ps:
     ShrHPSdict begin /ShrBORDERthickness 0 def}}
\def\mycorollary#1{\csname proclaim\endcsname{Corollary #1}}
\def\mythecorollary#1{\blue{#1}\immediate\special{ps:
     ShrHPSdict begin /ShrBORDERthickness 0 def}}
\def\mydefinition#1{\definition{Definition #1}}
\def\mythedefinition#1{\blue{#1}\immediate\special{ps:
     ShrHPSdict begin /ShrBORDERthickness 0 def}}
\def\myconjecture#1{\csname proclaim\endcsname{Conjecture #1}}
\def\myconjecturewithtitle#1#2{\csname proclaim\endcsname{Conjecture #1#2}}
\def\mytheconjecture#1{\blue{#1}\immediate\special{ps:
     ShrHPSdict begin /ShrBORDERthickness 0 def}}
\def\myproblem#1{\csname proclaim\endcsname{Problem #1}}
\def\myproblemwithtitle#1#2{\csname proclaim\endcsname{Problem #1#2}}
\def\mytheproblem#1{\blue{#1}\immediate\special{ps:
     ShrHPSdict begin /ShrBORDERthickness 0 def}}
\def\mytable#1{Table #1}
\def\mythetable#1{\blue{#1}\immediate\special{ps:
     ShrHPSdict begin /ShrBORDERthickness 0 def}}
\def\myanchortext#1#2{#2}
\def\mytheanchortext#1#2{\blue{#2}\immediate\special{ps:
     ShrHPSdict begin /ShrBORDERthickness 0 def}}
\font\eightcyr=wncyr8
\pagewidth{360pt}
\pageheight{606pt}
\topmatter
\title
Pseudo-Hadamard matrices of the first generation\\ and 
an algorithm for producing them.
\endtitle
\rightheadtext{Pseudo-Hadamard matrices of the first generation and \dots}
\author
Ruslan Sharipov
\endauthor
\address Bashkir State University, 32 Zaki Validi street, 450074 Ufa, Russia
\endaddress
\email
\myhref{mailto:r-sharipov\@mail.ru}{r-sharipov\@mail.ru}
\endemail
\abstract
Hadamard matrices in $\{0,1\}$ presentation are square $m\times m$ matrices 
whose entries are zeros and ones and whose rows considered as vectors 
in $\Bbb R^m$ produce the Gram matrix of a special form with respect to the 
standard scalar product in $\Bbb R^m$. The concept of Hadamard matrices is 
extended in the present paper. As a result pseudo-Hadamard matrices of the first 
generation are defined and investigated. An algorithm for generating these 
pseudo-Hadamard matrices is designed and is used for testing some conjectures. 
\endabstract
\endtopmatter
\loadbold
\TagsOnRight
\document

\head
1. Introduction.
\endhead
     Regular Hadamard matrices are defined in $\{1,-1\}$ presentation. They are 
square $n\times n$ matrices whose entries are ones and minus ones and whose rows 
are orthogonal to each other with respect to the standard scalar product in 
$\Bbb R^n$ (see \mycite{1}). Hadamard matrices are associated with Hadamard's 
maximal determinant problem (see \mycite{2} and \mycite{3}). A simplified version 
of this problem was suggested in \mycite{4}. Using the well-known transformation 
from $\{1,-1\}$ to $\{0,1\}$ presentation (see \mycite{2}), in \mycite{5} the concept
of Hadamard matrices was transferred to the class of matrices whose entries are zeros 
and ones. A Hadamard matrix in $\{0,1\}$ presentation is a special square $m\times m$ 
matrix, where $m=1$ or $m=4\,q-1$ for some $q\in\Bbb N$ and where $\Bbb N$ is the set 
of positive integers.\par
    The case $m=1$ is trivial. In this case we have exactly one Hadamard matrix
which coincides with the identity matrix: $H=\Vert 1\Vert$. In the case $m=4\,q-1$
Hadamard matrices in $\{0,1\}$ presentation can be defined as follows.
\mydefinition{1.1} A Hadamard matrix is a square $m\times m$ matrix, where $m=4\,q-1$
for some $q\in\Bbb N$, whose entries are zeros and ones and whose rows considered as 
vectors in $\Bbb R^m$ produce the Gram\footnotemark\ matrix of the form 
\footnotetext{\ A Gram matrix is a matrix formed by pairwise mutual scalar products 
of a sequence of vectors in a space equipped with some scalar product.}
\adjustfootnotemark{-1}
$$
\hskip -2em
G=\Vmatrix 
b & a & \hdots & a\\
a & b & \hdots & a\\
\vdots & \vdots & \ddots &\vdots\\
a & a & \hdots & b
\endVmatrix,
\mytag{1.1}
$$
where $a=q$ and $b=2\,q$, with respect to the standard scalar product in $\Bbb R^m$.
\enddefinition
    This definition is based on Theorems~2.1 and 2.2 from \mycite{5}. Below we
omit the case $m=1$ and consider the case $m=4\,q-1$ with $q\in\Bbb N$ only.\par
    Note that the restrictions $m=4\,q-1$, $a=q$, and $b=2\,q$ in
Definition~\mythedefinition{1.1} are essential since there is the following matrix
$$
M=\Vmatrix 
1 & 1 & 1 & 0\\
1 & 1 & 0 & 1\\
1 & 0 & 1 & 1\\
0 & 1 & 1 & 1
\endVmatrix\text{\ \ with \ }
G=\Vmatrix 
3 & 2 & 2 & 2\\
2 & 3 & 2 & 2\\
2 & 2 & 3 & 2\\
2 & 2 & 2 & 3
\endVmatrix
$$
which is not a Hadamard matrix, i\.\,e\. $M$ is not produced from a regular Hadamard
matrix by the $\{1,-1\}$ to $\{0,1\}$ transformation.\par 
\mytheorem{1.1} The transpose of a Hadamard matrix is again a Hadamard matrix.
\endproclaim
     Theorem~\mythetheorem{1.1} is immediate from Theorems 2.4 and 2.2 in \mycite{5}.
In particular this theorem means that the Gram matrix associated with columns of a
Hadamard matrix coincides with the Gram matrix \mythetag{1.1} associated with its
rows.\par
     Below by analogy to Definition~\mythedefinition{1.1} we define pseudo-Hadamard 
matrices in $\{0,1\}$ presentation and study a subclass of them. In particular, we 
design an algorithms for generating this subclass of pseudo-Hadamard matrices. 
\head
2. Pseudo-Hagamard matrices of the first generation.
\endhead
     Relying upon Definition~\mythedefinition{1.1} and Theorem~\mythetheorem{1.1}, it 
is easy to see that the set of Hadamard matrices is invariant under the following 
transformations:
\roster
\item"1)" permutation of rows; 
\item"2)" permutation of columns.
\endroster 
Using these transformations, one can bring any Hadamard matrix to the form
$$
\hskip -2em
H=\Vmatrix 1 & \hdots &1 & 0 & \hdots & 0\\
\vdots & \vtop{\hsize 75pt\vskip -12pt
\leftline{\vphantom{a}\kern -8pt
\boxed{\vtop to 57pt{\hsize 74pt
\vskip 22pt\centerline{$\{0,1\}$}
\vss}}}
\vskip -70pt\vss
}\kern -72pt\\
1\\
0\\
\vdots\\
0
\endVmatrix.
\mytag{2.1}
$$
Due to \mythetag{1.1} with $b=2\,q$ the number of ones in the first row of the
matrix \mythetag{2.1} is equal to $2\,q$. The number of zeros in this row is equal
to $2\,q-1$. Due to Theorem~\mythetheorem{1.1} the same is valid for the first
column of the matrix \mythetag{2.1}, i\.\,e\. its first column comprises $2\,q$
ones and $2\,q-1$ zeros.\par
     Let's remove the first row and the first column of the matrix in \mythetag{2.1}
and denote through $\tilde H$ the rest of the matrix $H$:
$$
\hskip -2em
\tilde H=\raise 31pt\hbox{
\boxed{\vtop to 57pt{\hsize 74pt
\vskip 22pt\centerline{$\{0,1\}$}
\vss}}}\ .
\mytag{2.2}
$$
The matrix \mythetag{2.2} coincides with the minor $M_{11}(H)$ in $H$ associated with
the top left entry of the matrix \mythetag{2.1}.\par
\mydefinition{2.1} An $m\times m$ matrix $\tilde H$, where $m=4\,q-2$ and $q\in\Bbb N$,
produced from some Hadamard matrix $H$ of the form \mythetag{2.1} according to 
\mythetag{2.2} is called a pseudo-Hadamard matrix of the first generation.
\enddefinition
\mytheorem{2.1} For any $m\times m$ pseudo-Hadamard matrix of the first generation
$\tilde H$ with $m=4\,q-2$, where $q\in\Bbb N$, its rows considered as vectors 
of\ \,$\Bbb R^m$ with the standard scalar product produce the Gram matrix of the form 
$$
\hskip -2em
\aligned
&\hphantom{aaaaa}\overbrace{\hphantom{aaaaaaaaaaa}}^{2\,q-1}
\hphantom{\,a}\overbrace{\hphantom{aaaaaaaaaaa}}^{2\,q-1}\\
\vspace{-1ex}
&G=\Vmatrix 
\tilde b & \tilde a & \hdots & \tilde a & a & a & \hdots & a\\
\tilde a & \tilde b & \hdots & \tilde a & a & a & \hdots & a\\
\vdots & \vdots & \ddots &\vdots & \vdots & \vdots & \ddots & \vdots\\
\tilde a & \tilde a & \hdots & \tilde b & a & a & \hdots & a\\
a & a & \hdots & a & b & a & \hdots & a\\
a & a & \hdots & a & a & b & \hdots & a\\
\vdots & \vdots & \ddots &\vdots & \vdots & \vdots & \ddots & \vdots\\
a & a & \hdots & a & a & a & \hdots & b\\
\endVmatrix,
\endaligned
\mytag{2.3}
$$
where $a=(m+2)/4=q$, $b=(m+2)/2=2\,q$, $\tilde a=a-1$, and $\tilde b=b-1$.
\endproclaim
\demo{Proof} Let's denote through $\tilde\bold r_1,\,\ldots,\,\tilde\bold r_m$ 
the rows of the matrix $\tilde H$ in \mythetag{2.2} and through $\bold r_0,\,\bold r_1,\,\ldots,\,\bold r_m$ the rows of the matrix $H$ in \mythetag{2.1}, i\.\,e\.
we denote the initial row of the matrix \mythetag{2.1} through $\bold r_0$. If we 
consider $\tilde\bold r_1,\,\ldots,\,\tilde\bold r_m$ as vectors in $\Bbb R^m$ and 
$\bold r_0,\,\bold r_1,\,\ldots,\,\bold r_m$ as vectors in $\Bbb R^{m+1}$, then, 
applying the standard scalar products in $\Bbb R^m$ and
$\Bbb R^{m+1}$ to them, we derive
$$
\hskip -2em
(\bold r_i,\bold r_j)=\sum^m_{k=0}H_{ik}\,H_{jk}=H_{i\,0}\,H_{j\,0}
+(\tilde\bold r_i,\tilde\bold r_j).
\mytag{2.4}
$$
Looking at \mythetag{2.1} and taking into account that $m=4\,q-2$, we see that
$$
\hskip -2em
H_{i\,0}=\cases 
1 & \text{for \ }0\leqslant i\leqslant 2\,q-1,\\
0 & \text{for \ }2\,q\leqslant i\leqslant m.
\endcases
\mytag{2.5}
$$
Since $a=(m+2)/4=q$ and $b=(m+2)/2=2\,q$, the formula \mythetag{1.1} 
is equivalent to 
$$
\hskip -2em
(\bold r_i,\bold r_j)=q\,(\delta_{ij}+1).
\mytag{2.6}
$$
Since moreover $\tilde a=a-1$ and $\tilde b=b-1$, the formula \mythetag{2.3} 
is equivalent to 
$$
\hskip -2em
\gathered
(\tilde\bold r_i,\tilde\bold r_j)=q\,(\delta_{ij}+1)-1\text{\ \ for \ }
1\leqslant i,j\leqslant 2\,q-1,\\
(\tilde\bold r_i,\tilde\bold r_j)=q\,(\delta_{ij}+1)\text{\ \ for \ }
2\,q\leqslant i\leqslant m\text{\ \ and/or \ }2\,q\leqslant j\leqslant m.
\endgathered
\mytag{2.7}
$$
Applying \mythetag{2.4} and \mythetag{2.5} to \mythetag{2.6}, we easily derive
\mythetag{2.7}. This means that \mythetag{2.6} implies \mythetag{2.7} and, hence,
\mythetag{1.1} implies \mythetag{2.3}. Theorem~\mythetheorem{2.1} is proved.
\qed\enddemo
     A similar result is valid for columns of pseudo-Hadamard matrices. It is
given by the following theorem. 
\mytheorem{2.2} For any $m\times m$ pseudo-Hadamard matrix of the first generation
$\tilde H$ with $m=4\,q-2$, where $q\in\Bbb N$, its columns considered as vectors 
of the space $\Bbb R^m$ with the standard scalar product produce the Gram matrix 
of the form \mythetag{2.3}, where $a=(m+2)/4=q$, $b=(m+2)/2=2\,q$, $\tilde a=a-1$, 
and $\tilde b=b-1$.
\endproclaim
     Theorem~\mythetheorem{2.2} follows from Theorem~\mythetheorem{2.1} due to
Theorem~\mythetheorem{1.1}. Theorems~\mythetheorem{2.1} and \mythetheorem{2.2}
are strengthened in the following theorem. 
\mytheorem{2.3} A square $m\times m$ matrix $\tilde H$ whose entries are zeros
and ones is a pseudo-Hadamard matrix of the first generation if and only if 
\,$m=4\,q-2$ for some $q\in\Bbb N$ and if its rows and its columns considered as 
vectors of the space $\Bbb R^m$ with the standard scalar product produce the 
same Gram matrix of the form \mythetag{2.3}, where $a=(m+2)/4=q$, $b=(m+2)/2=2\,q$, 
$\tilde a=a-1$, and $\tilde b=b-1$.
\endproclaim
\demo{Proof} The necessity part in the statement of Theorem~\mythetheorem{2.3} is 
proved by Theorems~\mythetheorem{2.1} and \mythetheorem{2.2}. Let's prove the 
sufficiency.\par
    In proving Theorems~\mythetheorem{2.1} we have seen that \mythetag{2.6} implies
\mythetag{2.7}. However the converse is not true since the equalities \mythetag{2.7} 
do not cover the cases with $i=0$ and $j=0$. Let's denote through $\tilde\bold r_0$
the initial row of the matrix \mythetag{2.1} shortened by omitting the first entry 
of it. This row obeys the equalities 
$$
\gather
\hskip -2em
(\tilde\bold r_0,\tilde\bold r_0)=2\,q-1,
\mytag{2.8}\\
\hskip -2em
(\tilde\bold r_0,\tilde\bold r_j)=q-1\text{\ \ for \ }
1\leqslant j\leqslant 2\,q-1,
\mytag{2.9}\\
\hskip -2em
(\tilde\bold r_0,\tilde\bold r_j)=q\text{\ \ for \ }
2\,q\leqslant j\leqslant m.
\mytag{2.10}
\endgather
$$\par
     The equalities \mythetag{2.8}, \mythetag{2.9}, and \mythetag{2.10} follow
from \mythetag{2.6} due to \mythetag{2.4} and \mythetag{2.5} and moreover, if we 
adjoin \mythetag{2.8}, \mythetag{2.9}, and \mythetag{2.10} to \mythetag{2.7}, 
the whole set of equalities \mythetag{2.7}, \mythetag{2.8}, \mythetag{2.9}, and 
\mythetag{2.10} turns out to be equivalent to \mythetag{2.6}. Therefore, in order
to complete our proof we need to derive \mythetag{2.8}, \mythetag{2.9}, and 
\mythetag{2.10} from the premises of Theorem~\mythetheorem{2.3} being proved.
\par
     The equality \mythetag{2.8} is trivial. It is fulfilled since the number of
ones in the row $\tilde\bold r_0$ is equal to $2\,q-1$. In order to derive 
\mythetag{2.9} and \mythetag{2.10} we define the following row:
$$
\hskip -2em
\aligned
&\hphantom{aaaa}\overbrace{\hphantom{aaaaaaaaaaa}}^{2\,q-1}
\hphantom{\,i}\overbrace{\hphantom{aaaaaaaaaaa}}^{2\,q-1}\\
\vspace{-1ex}
&\tilde\bold r=\Vmatrix 
1 & 1 & \hdots & 1 & 1 & 1 & \hdots & 1\\
\endVmatrix
\endaligned
\mytag{2.11}
$$
(compare with \mythetag{2.3}). The scalar products of the row \mythetag{2.11}
with $\tilde\bold r_0$ and with the rows of the matrix $\tilde H$ in the statement 
of Theorem~\mythetheorem{2.3} are easily calculated. They are equal to the number 
of ones in these rows:
$$
\gather
\hskip -2em
(\tilde\bold r,\tilde\bold r_0)=2\,q-1,\\
\hskip -2em
(\tilde\bold r,\tilde\bold r_j)=\tilde b=2\,q-1\text{\ \ for \ }
1\leqslant j\leqslant 2\,q-1,
\mytag{2.12}\\
\hskip -2em
(\tilde\bold r,\tilde\bold r_j)=b=2\,q\text{\ \ for \ }
2\,q\leqslant j\leqslant m.
\mytag{2.13}\\
\endgather
$$
Now let's consider the sum of all rows of the matrix $\tilde H$:
$$
\hskip -2em
\tilde{\boldsymbol\rho}=\sum^m_{i=1}\tilde\bold r_i.
\mytag{2.14}
$$
Each entry of the row \mythetag{2.14} is equal to the number of ones \pagebreak
in the corresponding column of the matrix $\tilde H$. Since $G$ in \mythetag{2.3}
is the Gram matrix not only for rows, but also for columns of the matrix $\tilde H$, 
the numbers of ones in columns of $\tilde H$ are given by diagonal entries 
of the Gram matrix \mythetag{2.3}. As a result we get
$$
\hskip -2em
\tilde{\boldsymbol\rho}=b\,\,\tilde\bold r+(\tilde b-b)\,\tilde\bold r_0
=2\,q\,\tilde\bold r-\tilde\bold r_0.
\mytag{2.15}
$$\par
     Let's calculate scalar products of both sides of \mythetag{2.15} with 
$\tilde\bold r_j$. In the case of the left hand side of \mythetag{2.15} we have 
the following result:
$$
\hskip -2em
(\tilde{\boldsymbol\rho},\tilde\bold r_j)=\sum^m_{i=1}(\tilde\bold r_i,\tilde\bold r_j)
=\sum^m_{i=1}G_{ij}.
\mytag{2.16}
$$
The last sum in \mythetag{2.16} is explicitly calculated using \mythetag{2.3}:
$$
\align
&\hskip -2em
\sum^m_{i=1}G_{ij}=(2\,q-2)\,(a+\tilde a)+a+\tilde b\text{\ \ for \ }
1\leqslant j\leqslant 2\,q-1,
\mytag{2.17}\\
&\hskip -2em
\sum^m_{i=1}G_{ij}=(2\,q-2)\,(a+a)+a+b\text{\ \ for \ }
2\,q\leqslant j\leqslant m.
\mytag{2.18}
\endalign
$$
Since $a=q$, $b=2\,q$, $\tilde a=a-1$, and $\tilde b=b-1$, \mythetag{2.17} and 
\mythetag{2.18} simplify to
$$
\hskip -2em
(\tilde{\boldsymbol\rho},\tilde\bold r_j)
=\sum^m_{i=1}G_{ij}=\cases 4\,q^2-3\,q+1 &\text{\ \ for \ }
1\leqslant j\leqslant 2\,q-1,\\
4\,q^2-q &\text{\ \ for \ }2\,q\leqslant j\leqslant m.
\endcases
\mytag{2.19}
$$\par
    Now let's proceed to the right hand side of \mythetag{2.15}. In this case
we have
$$
\gather
(2\,q\,\tilde\bold r-\tilde\bold r_0,\tilde\bold r_j)=2\,q\,(\tilde\bold r,
\tilde\bold r_j)-(\tilde\bold r_0,\tilde\bold r_j),
\mytag{2.20}\\
\vspace{-3ex}
\intertext{where}
\vspace{-3ex}
(\tilde\bold r,\tilde\bold r_j)=\cases
\tilde b &\text{\ \ for \ }
1\leqslant j\leqslant 2\,q-1,\\
b &\text{\ \ for \ }2\,q\leqslant j\leqslant m
\endcases
\mytag{2.21}
\endgather
$$
(see \mythetag{2.12} and  \mythetag{2.13}). Since $b=2\,q$ and $\tilde b=b-1$, 
from \mythetag{2.20} and \mythetag{2.21} we get 
$$
\hskip -3em
(2\,q\,\tilde\bold r-\tilde\bold r_0,\tilde\bold r_j)=-(\tilde\bold r_0,
\tilde\bold r_j)+\cases
4\,q^2-2\,q &\text{\ \ for \ }
1\leqslant j\leqslant 2\,q-1,\\
4\,q^2 &\text{\ \ for \ }2\,q\leqslant j\leqslant m.
\endcases
\hskip -1em
\mytag{2.22}
$$\par
    Note that \mythetag{2.15} implies $(\tilde{\boldsymbol\rho},\tilde\bold r_j)
=(2\,q\,\tilde\bold r-\tilde\bold r_0,\tilde\bold r_j)$. Substituting 
\mythetag{2.19} and \mythetag{2.22} into this equality, we get two expressions for
$(\tilde\bold r_0,\tilde\bold r_j)$:
$$
\align
&\hskip -2em
(\tilde\bold r_0,\tilde\bold r_j)=q-1\text{\ \ for \ }
1\leqslant j\leqslant 2\,q-1,
\mytag{2.23}\\
&\hskip -2em
(\tilde\bold r_0,\tilde\bold r_j)=q\text{\ \ for \ }
2\,q\leqslant j\leqslant m.
\mytag{2.24}
\endalign    
$$
Note that \mythetag{2.23} and  \mythetag{2.24} do coincide with 
\mythetag{2.9} and  \mythetag{2.10}.\par
     Thus, the formulas \mythetag{2.8}, \mythetag{2.9}, and \mythetag{2.10}
are derived from the premises of Theorem~\mythetheorem{2.3}. Adjoining them
to \mythetag{2.7} and applying \mythetag{2.4} and \mythetag{2.5}, we derive 
\mythetag{2.6} for the rows of the matrix $H$. The matrix $H$ now is produced
backward from $\tilde H$ by adjoining the initial row and the initial column
according to \mythetag{2.1} and  \mythetag{2.2}. \pagebreak The equality 
\mythetag{2.6} is equivalent to \mythetag{1.1}. Therefore we can apply either 
Theorem~2.2 or Theorem~2.4 from \mycite{5}. Each of these two theorems means 
that the matrix $H$ produced backward from $\tilde H$ in \mythetag{2.1} is a 
regular Hadamard matrix in $\{0,1\}$ presentation. Hence $\tilde H$ is a 
pseudo-Hadamard matrix of the first generation according to 
Definition~\mythedefinition{1.1}. The proof of Theorem~\mythetheorem{2.3} 
is over.
\qed\enddemo
\head
3. Pseudo-Hadamard matrices of higher generations.
\endhead
     Pseudo-Hadamard matrices of higher generations are defined recursively.
A pseu\-do-Hadamard matrix of the second generation is produced from some 
pseudo-Hadamard matrix of the first generation upon rearranging its rows and columns 
in a way similar to \mythetag{2.1} and then by removing the initial row and the initial 
column of it like in \mythetag{2.2}. Matrices of the third generation are produced 
in this way from matrices of the second generation etc, i\.\,e\. each next generation 
is produced from the previous one.\par
     In this paper we shall not consider pseudo-Hadamard matrices of higher generations.
They will be studied separately in forthcoming papers.\par 
\head
4. An algorithm for generating pseudo-Hadamard matrices of the first
generation.
\endhead
    Like the algorithm for generating Hadamard matrices from \mycite{5}, our present
algorithm is based on partitioning of rows of matrices into groups (see Section 3 in
\mycite{5}). We use the Maxima programming language (see \mycite{6}) for presenting
its code. Almost all of the code coincide with the code in \mycite{5}. Below are
those lines of the code that should be changed. 
\medskip
\parshape 1 10pt 350pt 
{\tt\noindent\darkred{HM\_size:m\$}\newline
\darkred{HM\_quarter:(HM\_size+1)/4\$\kern -12.5em
\raise 3pt\hbox to 14em{\vbox{\hrule width 13em height 0.5pt}\hss}}\newline
\green{HM\_quarter:(HM\_size+2)/4\$}\newline
\darkred{q:HM\_quarter\$}\newline
\darkred{HM\_row[1]:[[0,2*q],[1,2*q-1]]\$\kern -15.2em
\raise 3pt\hbox to 17em{\vbox{\hrule width 15.7em height 0.5pt}\hss}}\newline
\green{HM\_row[1]:[[0,2*q-1],[1,2*q-1]]\$}\newline
\darkred{HM\_row[2]:[[0,q],[1,q],[2,q],[3,q-1]]\$\kern -19.2em
\raise 3pt\hbox to 20em{\vbox{\hrule width 19.7em height 0.5pt}\hss}}\newline
\green{HM\_row[2]:[[0,q-1],[1,q],[2,q],[3,q-1]]\$}\newline
\darkred{HM\_matrix\_num:1\$}\newline
\darkred{HM\_stream:openw("output\_file.txt")\$}\newline
\darkred{HM\_make\_row(3)\$}\newline
\darkred{close(HM\_stream)\$}
}
\medskip
\noindent
The lines to be removed are shown with strikethrough text. The replacement lines 
are given in green. Like in \mycite{5}, the whole job is practically done 
by the recursive function {\tt\darkred{HM\_make\_row()}}. Here its code is also 
slightly changed. 
\medskip
\parshape 1 10pt 350pt 
{\tt\noindent\darkred{HM\_make\_row(i):=block}\newline
\darkred{\ ([n,s,k,l,q,dummy,kk,y,dpnd,indp,nrd,nri,r,kr,qq,eq,eq\_list,j,}\newline
\darkred{\ \ LLL,RLL,RVV,RRV,subst\_list],}\newline
\darkred{\ \ if not integerp(HM\_size) or HM\_size<3 or mod(HM\_size,4)\#3\kern -28.5em
\raise 3pt\hbox to 29em{\vbox{\hrule width 28.7em height 0.5pt}\hss}}\newline
\green{\ \ if not integerp(HM\_size) or HM\_size<2 or mod(HM\_size,4)\#2}\newline
\darkred{\ \ \ then}\newline
\darkred{\ \ \ \ (}\newline
\darkred{\ \ \ \ \ print(printf(false,"Error:\ m=\~{}a is incorrect size for }\newline
\darkred{\ \ \ \ \ Hadamard matrices",HM\_size)),}\newline
\green{\ \ \ \ \ generation one pseudo-Hadamard matrices",HM\_size)),}\newline
\darkred{\ \ \ \ \ return(false)}\newline
\darkred{\ \ \ \ ),}\newline
\darkred{\ \ if HM\_size=3\kern -6.2em
\raise 3pt\hbox to 7em{\vbox{\hrule width 6.5em height 0.5pt}\hss}}\newline
\green{\ \ if HM\_size=2}\newline
\darkred{\ \ \ then}\newline
\darkred{\ \ \ \ (}\newline
\darkred{\ \ \ \ \ HM\_row[2]:[[0,1],[1,1],[2,1]],\kern -15.2em
\raise 3pt\hbox to 16em{\vbox{\hrule width 15.5em height 0.5pt}\hss}}\newline
\green{\ \ \ \ \ HM\_row[2]:[[1,1],[2,1]],}\newline
\darkred{\ \ \ \ \ HM\_row[3]:[[1,1],[2,1],[4,1]],\kern -15.2em
\raise 3pt\hbox to 16em{\vbox{\hrule width 15.5em height 0.5pt}\hss}}\newline
\darkred{\ \ \ \ \ HM\_output\_matrix(),}\newline
\darkred{\ \ \ \ \ return(false)}\newline
\darkred{\ \ \ \ ),}\newline
\darkred{\ \ print(printf(false,"i=\~{}a",i)),}\newline
\darkred{\ \ ..............................}\newline
\darkblue{\ \ /*-- prepare the equation list --*/}\newline
\darkred{\ \ eq\_list:[],}\newline
\darkred{\ \ var\_list:[],}\newline
\darkred{\ \ ..............................}\newline
\darkred{\ \ eq\_list:endcons(eq=2*HM\_quarter,eq\_list),\kern -20.5em
\raise 3pt\hbox to 21em{\vbox{\hrule width 20.7em height 0.5pt}\hss}}\newline
\green{\ \ if i<2*HM\_quarter}\newline
\green{\ \ \ then eq\_list:endcons(eq=2*HM\_quarter-1,eq\_list)}\newline
\green{\ \ \ else eq\_list:endcons(eq=2*HM\_quarter,eq\_list),}\newline
\darkred{\ \ qq:1,}\newline
\darkred{\ \ ..............................}\newline
\darkred{\ \ \ \ eq\_list:endcons(eq=HM\_quarter,eq\_list),\kern -19.5em
\raise 3pt\hbox to 20em{\vbox{\hrule width 19.7em height 0.5pt}\hss}}\newline
\green{\ \ \ \ if i<2*HM\_quarter}\newline
\green{\ \ \ \ \ then eq\_list:endcons(eq=HM\_quarter-1,eq\_list)}\newline
\green{\ \ \ \ \ else eq\_list:endcons(eq=HM\_quarter,eq\_list),}\newline
\darkred{\ \ \ \ qq:qq*2}\newline
\darkred{\ \ ..............................}
}
\smallskip
\noindent
For the sake of brevity above we omit some unchanged portions of the code replacing
them with dots. The lacking code can be taken from \mycite{5}.\par
     Apart from the function {\tt\darkred{HM\_make\_row()}} the algorithm 
comprises two other functions {\tt\darkred{HM\_output\_matrix()}} and 
{\tt\darkred{HM\_sc\_prods\_ok(i)}}. Their code is unchanged. It can also be 
taken from \mycite{5}.\par
     Like in \mycite{5}, the above code was run in Maxima, version 5.42.2, on Linux 
platform of Ubuntu 16.04 LTS using laptop computer DEXP Atlas H161 with the processor
unit Intel Core i7-4710MQ. Below are performance data of the code.\par
{\bf The case $m=2$} is trivial. In this case the algorithm terminated instantly and 
produced exactly one $2\times 2$ pseudo-Hadamard matrix which coincides with the
identity matrix.\par
{\bf The case $m=6$} is less trivial. In this case the algorithm also
terminated instantly, but produced 6 matrices.\par
{\bf The case $m=10$}. In this case the algorithm ran for 6 seconds and produced
1440 matrices. The matrix production rate is 14400 matrices/minute.\par
{\bf The case $m=14$}. In this case the algorithm did not terminate during
observably short time. But setting timestamps upon each next 10000 matrices, I
have found that the first 10000 matrices were produced for 56 seconds, i\.\,e\.
the matrix production rate is 10714 matrices/minute.\par
{\bf The case $m=18$}. In this case the first 10000 matrices were produced for 1 minute 
and 29 seconds, i\.\,e\. the matrix production rate is 6742 matrices/minute.\par
{\bf The case $m=22$} is different. In this case the algorithm becomes
very slow. It has produced 10000 matrices upon running for 5 hours 1 minute and 17 seconds.
The average matrix production rate is 33 matrices/minute. However this production rate 
is very unevenly distributed over the interval of running. In the beginning the algorithm
does not produce matrices for about 3 hours.\par
     As a conclusion we can say that $m=22$ is a practical limit for the algorithm in its
present version. Though theoretically the algorithm has no limits. 
\head
5. Analysis of output and some conjectures.
\endhead
     The above algorithm produces $m\times m$ matrices whose entries are zeros and ones
and whose rows, when treated as vectors in $\Bbb R^m$, generate the Gram matrix of the form
\mythetag{2.3} with respect to the standard scalar product in $\Bbb R^m$. However we cannot
apply Theorem~\mythetheorem{2.3} to these matrices since the Gramians of their columns are
uncertain. Therefore the output matrices were additionally analyzed. Relying upon this
analysis the following conjectures are formulated. 
\myconjecture{5.1} Let $\tilde H$ be a square $m\times m$ matrix, where $m=4\,q-2$ for 
some $q\in\Bbb N$, whose entries are zeros and ones and whose rows considered as 
vectors of the space $\Bbb R^m$ with the standard scalar product produce the Gram matrix 
of the form \mythetag{2.3} with $a=(m+2)/4=q$, $b=(m+2)/2=2\,q$, $\tilde a=a-1$, and $\tilde b=b-1$. Then $\tilde H$ coincides with some  pseudo-Hadamard matrix of the first generation upon some permutation of its columns.
\endproclaim
\myconjecture{5.2} Let $\tilde H$ be a square $m\times m$ matrix, where $m=4\,q-2$ for 
some $q\in\Bbb N$, whose entries are zeros and ones and whose columns considered as 
vectors of the space $\Bbb R^m$ with the standard scalar product produce the Gram matrix 
of the form \mythetag{2.3} with $a=(m+2)/4=q$, $b=(m+2)/2=2\,q$, $\tilde a=a-1$, and $\tilde b=b-1$. Then $\tilde H$ coincides with some  pseudo-Hadamard matrix of the first generation upon some permutation of its rows.
\endproclaim
Conjectures~\mytheconjecture{5.1} and \mytheconjecture{5.2} are dual to each other. They
are either both valid or both invalid. I have tested these conjectures for all of my
output matrices. They turned out to be valid
\roster
\item"1)" for six $m\times m$ matrices with $m=6$;
\item"2)" for one thousand four hundred and forty $10\times 10$ matrices;
\item"3)" for ten thousand $14\times 14$ matrices;
\item"4)" for ten thousand $18\times 18$ matrices;
\item"5)" for ten thousand $22\times 22$ matrices.
\endroster
This makes a good evidence in favor of these conjectures to be valid, though this does not
prove them.\par
\head
6. Dedicatory.
\endhead
     This paper is dedicated to my sister Svetlana Abdulovna Sharipova. 

\enddocument
\end